\documentstyle[twoside,fleqn,espcrc2,epsf]{article}

\title{Lattice QCD at finite isospin density.}

\author{J.~B.~Kogut\address{Physics Department, University of Illinois,
        1110 West Green Street, Urbana, IL 61801, USA},
        D.~K.~Sinclair\address{HEP Division, Argonne National Laboratory,
        9700 South Cass Avenue, Argonne, IL 60439, USA}
        \thanks{Talk presented by D.~K.~Sinclair at Lattice2001, Berlin, 
                19th--24th August, 2001.}}

\begin{document}

\begin{abstract}
We have simulated QCD at a finite chemical potential $\mu_I$ for isospin ($I_3$)
to probe part of the phase diagram for nuclear matter. Preliminary results
suggest that for $\mu_I > \mu_c$, this theory forms a charged pion condensate
which spontaneously breaks $I_3$, and the isospin density is non zero.
\vspace{1pc}
\end{abstract}

\maketitle

\section{Introduction}

Nuclear matter has finite baryon number density, finite (negative) isospin
density and, under certain conditions, finite (negative) strangeness density.
Since having a finite chemical potential for baryon number makes the fermion
determinant complex and simulations intractable, we restrict ourselves to zero
baryon-number chemical potential. QCD at finite chemical potential $\mu_I$ for
isospin ($I_3$) and zero chemical potential for baryon number, has a real
positive fermion determinant and can thus be simulated. This describes a
surface in the phase diagram for nuclear matter. This theory turns out to be
very similar to 2-colour QCD at finite quark-number density which we are also
studying \cite{kshm,kts}. We will also use various devices to include the
effects of finite chemical potential $\mu_S$ for strangeness.

QCD at finite chemical potential $\mu_I$ for isospin has been studied by Son
and Stephanov \cite{ss} using effective (chiral) lagrangians. This
analysis indicates is that it undergoes a transition at $|\mu_I|=\mu_c=m_\pi$
to a state with a charged pion condensate, while the orthogonal charged pion
becomes a true Goldstone boson. This condensate breaks $I_3$ and parity
spontaneously.

Section 2 describes our simulations of 2 flavour QCD with a finite $\mu_I$.
A brief discussion of how to include a finite $\mu_S$ is included. Section 3
gives our conclusions.

\section{Lattice QCD at finite isospin chemical potential}

The lattice fermion action (staggered quarks) for QCD with isospin chemical
potential, $\mu_I$ is
\begin{equation}
S_f=\sum_{sites} \left[\bar{\chi}[D\!\!\!\!/(\tau_3\mu_I)+m]\chi
                   + i\lambda\epsilon\bar{\chi}\tau_2\chi\right]
\end{equation}
where $D\!\!\!\!/(\mu)$ is the standard staggered $D\!\!\!\!/$ with links in
the $+t$ direction multiplied by $e^{\frac{1}{2}\mu}$ and those in the $-t$
direction multiplied by $e^{-\frac{1}{2}\mu}$. The term proportional to
$\lambda$, which explicitly breaks the third component of isospin, $I_3$, is
introduced to allow us to observe spontaneous breaking of this symmetry from
finite lattice simulations. We will be interested in the limit
$\lambda \rightarrow 0$. The Dirac operator is
\begin{equation}
{\cal M}=\left[\begin{array}{cc}
               D\!\!\!\!/(\mu_I)+m         &           \lambda\epsilon       \\
               -\lambda\epsilon            &           D\!\!\!\!/(-\mu_I)+m
               \end{array}                                            \right]
\end{equation}
whose determinant
\begin{equation}
\det{\cal M}=\det[{\cal A}^\dagger{\cal A}+\lambda^2]
\end{equation}
where
\begin{equation}
{\cal A} \equiv D\!\!\!\!/(\mu_I)+m.
\end{equation}
Since this determinant is strictly positive, standard simulation methods work.
We use hybrid molecular dynamics simulations to tune the number of flavours
down from 8 to 2.   

We measure the chiral condensate 
\begin{equation}
\langle\bar{\psi}\psi\rangle \Leftrightarrow \langle\bar{\chi}\chi\rangle,
\end{equation}
the isospin condensate
\begin{equation}
i\langle\bar{\psi}\gamma_5\tau_2\psi\rangle  \Leftrightarrow 
i\langle\bar{\chi}\epsilon\tau_2\chi\rangle
\end{equation}
and the isospin density
\begin{equation}
j_0^3 = \left\langle{\partial S_f \over \partial\mu_I}\right\rangle
\end{equation}

At $\mu_I=\lambda=0$, there are 4 pseudo-Goldstone pions corresponding to
\begin{equation}
i\langle\bar{\chi}\epsilon\tau_i\chi\rangle
\end{equation}
\begin{equation}
i\langle\bar{\chi}\epsilon\chi\rangle
\end{equation}
Only 2 of these (those with $i=1,2$) correspond to pseudo-Goldstone pions of
the $N_f=2$ theory --- the third $N_f=2$ pion is a massive 3-link object at
finite lattice spacing. However, the other 2 Goldstone bosons should mimic the
behaviour of the third $N_f=2$ pion ($\pi_0$) in the continuum limit and are
thus useful to consider. As $\mu_I$ increases one expects the $i=1$ mass to
fall, while the $i=2$ mass should rise. The other 2 masses are expected to 
remain unchanged for small $\mu_I$. One expects a phase transition at
$\mu_I=\mu_c\approx m_\pi$, above which isospin($I_3$) is spontaneously broken
and the $i=1$ state becomes a true Goldstone boson. This is only part of the
story. Once $j_0^3$ gains an expectation value, scalar mesons are also 
candidate pseudo-Goldstone bosons.

We are simulating this theory at $\beta=5.2$, $m=0.025,0.05$,
$\lambda=0.005,0.0025$ on an $8^4$ lattice. We are also performing simulations
at $m=0.05$, $\lambda=0.005$ on an $8^3 \times 4$ lattice to study the finite
temperature transition. 

\begin{figure}[htb]
\epsfxsize=3in
\centerline{\epsffile{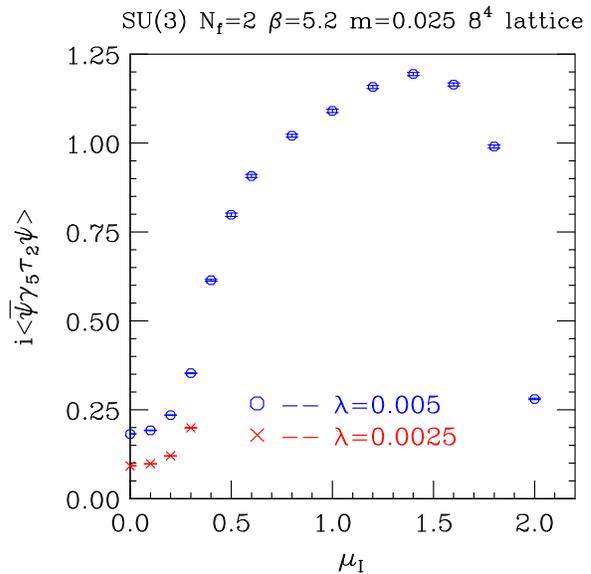}}
\caption{The pion condensate as a function of $\mu_I$.}\label{fig:pbg5t2p}
\end{figure}

The pion condensate is shown in figure~\ref{fig:pbg5t2p}
This strongly suggests that 
there is some $\mu_I = \mu_c$ above which isospin($I_3$) is spontaneously
broken by a charged pion condensate. However, we will need to finish our
simulations at $\lambda=0.0025$ and extrapolate to $\lambda=0$ to validate
this observation.

\begin{figure}[htb]      
\epsfxsize=3in   
\centerline{\epsffile{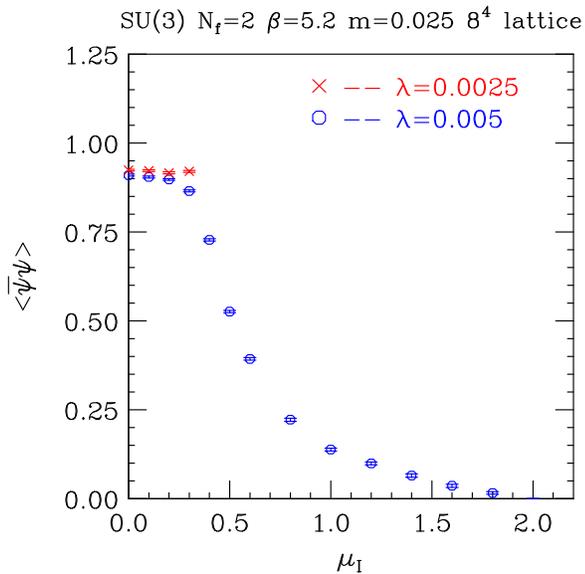}}
\caption{The chiral condensate as a function of $\mu_I$.}\label{fig:pbp}
\end{figure}

The chiral condensate shown in figure~\ref{fig:pbp}
is approximately constant for $\mu_I < \mu_c$.
For $\mu_I > \mu_c$ it starts to fall, indicating that the condensate is
rotating from the chiral to the isospin-breaking direction --- it does
not appear to be a simple rotation which is suggested by the tree-level
effective Lagrangian approach, however.

\begin{figure}[htb]
\epsfxsize=3in   
\centerline{\epsffile{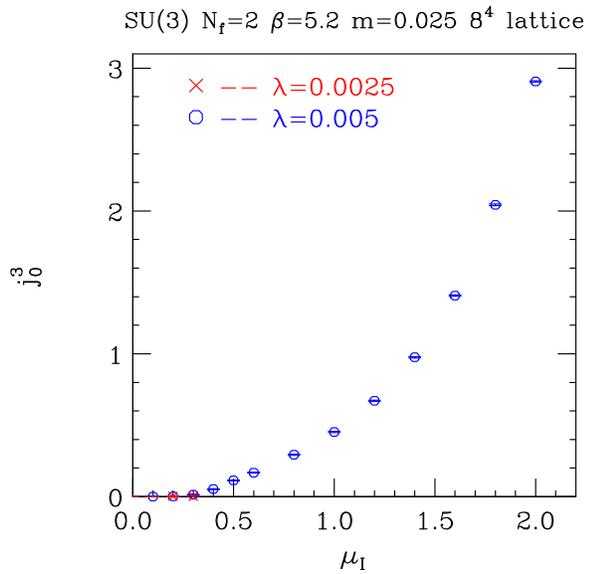}}
\caption{The density of the 3rd component of isospin as a function of $\mu_I$.}
\label{fig:j0} 
\end{figure}

The isospin density, shown in figure~\ref{fig:j0} is consistent with zero for 
$\mu_I < \mu_c$ and commences to rise for $\mu_I > \mu_c$. It reaches its
saturation value (3) for $\mu_I \sim 2$.

We will measure the spectrum of Goldstone and pseudo-Goldstone mesons on a
larger ($12^3 \times 24$) lattice. In addition we will store configurations
for offline calculations of other hadron spectra ($N$, $\Delta$, $\rho$...)
which are also expected to exhibit interesting behaviour.

\section{Conclusions}

QCD at finite chemical potential for isospin($I_3$) appears to spontaneously
break isospin with a charged pion condensate, for $\mu_I$ large enough. We need
another $\lambda$ to verify this and to show that $\mu_c > 0$. We will repeat
these simulations on a larger ($12^3 \times 24$) lattice to measure the
spectrum, in particular the Goldstone boson. Measurement of the instanton
distribution is of interest. Is there any additional structure at high $\mu_I$,
where the effective Lagrangian approach is suspect, perhaps related to baryon
thresholds? What we really would like to know is do such pion condensates form
at finite baryon-number density (i.e. in nuclear matter) for $\mu_I$ large 
enough?

Preliminary indications from our finite temperature ($8^3 \times 4$)
simulations are that the evaporation of the pion condensate for large enough
temperature occurs at a first order transition. This needs further study.

Simulations with finite chemical potentials for isospin and strangeness will
study the competition between pion and kaon condensation \cite{kt}
Two methods will be used to avoid a complex determinant. The first is to use a
partially quenched approach where only the $u$ and $d$ quarks are dynamical.
The second is to include a $c$ quark with $m_c = m_s$ and a chemical potential
for the 3rd component of the $SU(2)$ flavour group which mixes $c$ and $s$.

QCD at finite isospin, and QCD at finite isospin and strangeness should give
sensible quenched results. In fact our earlier work on quenched QCD at finite
baryon number density \cite{kls} can be reinterpreted as quenched QCD at
finite isospin density.

\section*{Acknowledgements}

DKS was supported by DOE contract W-31-109-ENG-38. JBK was supported in part
by an NSF grant NSF-PHY96-05199. These simulations were performed on the Cray
SV1's and IBM SP's at NERSC and the Cray T90 and IBM SP's at NPACI.

\end{document}